\documentclass[conference]{IEEEtran}
\IEEEoverridecommandlockouts
\usepackage{cite}
\usepackage{amsmath,amssymb,amsfonts}
\usepackage{algorithmic}
\usepackage{graphicx}
\usepackage{textcomp}
\usepackage{xcolor}

\newcommand{\sectopic}[1]{\vspace{0.2em}\par\noindent{\textit{\bfseries #1}}}

\newcommand{\projname}{\textit{CompliAT}}

\def\BibTeX{{\rm B\kern-.05em{\sc i\kern-.025em b}\kern-.08em
    T\kern-.1667em\lower.7ex\hbox{E}\kern-.125emX}}
\begin{document}

\title{Towards Standards-Compliant Assistive Technology Product Specifications via LLMs\\
}

\author{\IEEEauthorblockN{Chetan Arora, John Grundy}
\IEEEauthorblockA{\textit{Faculty of IT} \\
\textit{Monash University}\\
Melbourne, Australia \\
\{chetan.arora, john.grundy\}@monash.edu}
\and
\IEEEauthorblockN{Louise Puli, Natasha Layton}
\IEEEauthorblockA{\textit{Rehabilitation, Ageing and Independent Living} \\ \textit{Research Centre (RAIL), Monash University} \\
Melbourne, Australia \\
\{louise.puli, natasha.layton\}@monash.edu}
}

\maketitle

\begin{abstract}
In the rapidly evolving field of assistive technology~(AT), ensuring that products meet national and international standards is essential for user safety, efficacy, and accessibility. In this vision paper, we introduce \projname, a pioneering framework designed to streamline the compliance process of AT product specifications with these standards through the innovative use of Large Language Models (LLMs). \projname~addresses three critical tasks: checking terminology consistency, classifying products according to standards, and tracing key product specifications to standard requirements.
We tackle the challenge of terminology consistency to ensure that the language used in product specifications aligns with relevant standards, reducing misunderstandings and non-compliance risks.
We propose a novel approach for product classification, leveraging a retrieval-augmented generation model to accurately categorize AT products aligning to international standards, despite the sparse availability of training data. Finally, \projname~implements a traceability and compliance mechanism from key product specifications to standard requirements, ensuring all aspects of an AT product are thoroughly vetted against the standards.
By semi-automating these processes, \projname~aims to significantly reduce the time and effort required for AT product standards compliance and uphold quality and safety standards. We outline our planned implementation and evaluation for \projname. 
\end{abstract}


\section{Introduction}~\label{sec:introduction}
Assistive technology (AT) refers to devices or systems that help individuals with disabilities perform tasks they might otherwise find difficult or impossible, such as hearing aids for those with hearing impairments. In the AT sector, the requirements and product specifications must comply with national and international standards for activities from initial product design and development to post-market surveillance and reporting. Ensuring compliance with these standards is not just about adhering to regulatory mandates; it is essential for guaranteeing the safety, efficacy, and accessibility of these technologies for the diverse populations they serve~\cite{cooper1998harmonization}. Compliance with relevant standards is fundamental to the trust users place in AT products, the confidence healthcare providers have in recommending them, and providing legislative rulings for their usage. 

The AT sector continues to grow, driven by technological advancements and an increasing recognition of the needs of ageing populations and individuals with disabilities. This means the challenge of maintaining AT product compliance has become increasingly more important but also more difficult. \emph{Standards} serve as vital benchmarks that guide the development and deployment of AT products, ensuring they meet universal criteria for quality and functionality. The complexity of AT products and the evolving nature of international standards demands a more structured and semi-automated approach is needed to ensure AT device compliance. We outline a novel idea for building a comprehensive framework -- (\projname) -- to streamline AT product specifications and requirements compliance through three pivotal tasks. This aims at better aligning AT products with crucial benchmarks using advanced natural language processing (NLP) techniques, particularly large language models (LLMs). We briefly discuss the three key tasks of \projname~below.

\sectopic{1. Terminology Consistency Checking:} This ensures the terminology used in product specifications aligns with those prescribed in their relevant standards. Given the technical specificity of AT products and the precision required in standards, inconsistencies in terminology can lead to misunderstandings, misinterpretations, and, ultimately, non-compliance. For example, according to ISO standard ISO9999:2022~\cite{iso9999} on the classification and terminology of AT products, ``Rollators or wheelie walkers'' refers to a specific product type for mobility support. In the product market, it is often referred to as a buggy, shopper walker, or gait trainer, often leading to confusion and misinterpretation regarding its classification. This discrepancy in terminology can result in manufacturers inadvertently failing to meet essential criteria outlined by relevant standards. A detailed analysis of the product specifications and the standards to identify and resolve any discrepancies in terminology is required.

\sectopic{2. Product Classification According to Standards:} This focuses on correctly classifying AT products according to the categories defined by the standards. Classification is pivotal because it determines the specific standards applicable to each product. Misclassification can result in overlooking critical compliance requirements, adhering to irrelevant ones, or simply not reaching the right user base, compromising the product's compliance status. For example, in product specifications, if a product is referred to as a `shopper walker', it should be correctly classified according to the classes in ISO999:2022 to ``12 06 06 
Rollators or wheelie walkers'', where 12 06 06 is the unique class code for this product category. Understanding the standards' classification system and analyzing various product features to ensure accurate classification is required.

\sectopic{3. Compliance Checking Against Product Categories and Standards:} This involves thoroughly comparing AT product specifications against the requirements of the correct product categories and other cross-referenced standards. Comprehensive compliance checking ensures that every aspect of the product, from design and functionality to safety and efficacy, meets the requirements. 
A detailed understanding of the product specifications and the requirements of the applicable standards is required.

These tasks are the foundation of our \projname~framework, ensuring that AT product specifications comply with their relevant standards. By systematically addressing each task, our framework aims to streamline the compliance process, reduce the risk of non-compliance, and ultimately enhance the quality and reliability of AT products. No comprehensive solution is available for compliance support to manufacturers, policymakers, and standards' decision makers to facilitate compliance with AT product specifications to standards~\cite{layton2012barriers,unsworth2021powered}. Hence, several key national organizations across the globe fail to adhere to relevant AT standards~\cite{layton2023time}. We aim to address this major gap in AT research and industry and present our vision of \projname~framework in this paper. We posit that, if accepted, the REWBAH workshop will be an excellent venue for discussing our vision in this paper and receiving feedback from RE experts on this novel topic.

\sectopic{Structure.} Section~\ref{sec:Example} presents a motivating example for the three key tasks in \projname. Section~\ref{sec:approach} discusses our vision of the \projname~framework and the implementation plan. 
Section~\ref{sec:related} covers the related work and Section~\ref{sec:conclusion} concludes the paper.

\section{Motivating Example}~\label{sec:Example}
Consider a hypothetical scenario involving the National Disability Management Department (NDMD) of the state of ``Futuria''. NDMD oversees and facilitates the distribution and utilization of digital and smart assistive technology (AT) products, e.g., AT software solutions, automated wheelchairs, and smart canes, across the healthcare and social services sectors, ensuring that these products meet the needs of individuals with disabilities effectively and comply with international standards for tasks, such as the three tasks discussed earlier. This facilitates a standardized approach to product development and ensures that these innovations are readily accepted and integrated into the healthcare and social services sectors.

\sectopic{Smart Knee AT specification.} Consider a new microprocessor-controlled `smart' prosthetic knee (StrideTech-ProKnee) AT that offers amputees mobility and stability. StrideTech-ProKnee incorporates a microprocessor that continuously analyzes data from sensors monitoring the user's movement and the surrounding environment in real-time. It adjusts the knee's internal mechanical resistance and movement to mimic natural gait patterns as closely as possible. One safety feature is a built-in fall detector, designed to predict and react to conditions that might lead to a fall, significantly reducing the risk of injury. This fall detector can preemptively adjust the knee's behaviour to provide stability when a potential fall is detected, such as locking the knee or adjusting its swing to help the user regain their balance. It also has the potential to alert others should the user fall and need help, managed via the StrideTech-ProKnee app. StrideTech-ProKnee can also collect data on near-miss falls and actual falls, which can be used to guide prosthetic knee prescription and suitability. The knee connects directly to the user's smartphone or iPad via the app. 

\sectopic{Smart Knee AT's product specification compliance.} StrideTech-ProKnee's product specification needs to be cross-referenced by NDMD regarding terminological consistency and vocabulary usage with the ISO9999 Assistive products vocabulary. Further, NDMD requires correct classification of such products, e.g., to categorize StrideTech-ProKnee based on their primary function under categories ``06 Orthoses and prostheses $>$ 06 24 Lower limb prostheses'' and
``06 Orthoses and prostheses $>$ 06 24 33 Knee units'', and based on their secondary functions into categories, e.g., ``22 Assistive products that record, play and display audio and visual information'', 
``22 29 Assistive products for signaling, alarming and localization'', and 
``22 29 06 Personal emergency alarm systems''. Based on these categorizations, StrideTech-ProKnee product specifications will also need to be checked for compliance against related prosthetics standards, such as ISO 8549-1:2020~\cite{ISO8549} for prosthetics and orthotics Vocabulary and
ISO 10328:2016~\cite{ISO10328} prosthetics structural testing of lower-limb prostheses. ISO 10328:2016 involves ensuring that the product specification and the manufacturing documentation adhere to internationally recognized benchmarks for safety and performance. For example, for StrideTech-ProKnee specification \textit{``The StrideTech-ProKnee is equipped with an advanced sensor array and microprocessor system designed to adjust knee resistance and movement in real-time, optimizing user mobility and stability. This system includes a feature for automatic lock adjustment under varying load conditions, ensuring user safety and comfort.''}, it should be traced to the Proof of Strength testing mechanism and recommend test scenarios for ``Structure shall sustain static loading by proof test forces at prescribed values for prescribed times'' in ISO 10328:2016.  

\begin{figure*} [!t]
\centering
\includegraphics[width=0.75\textwidth]{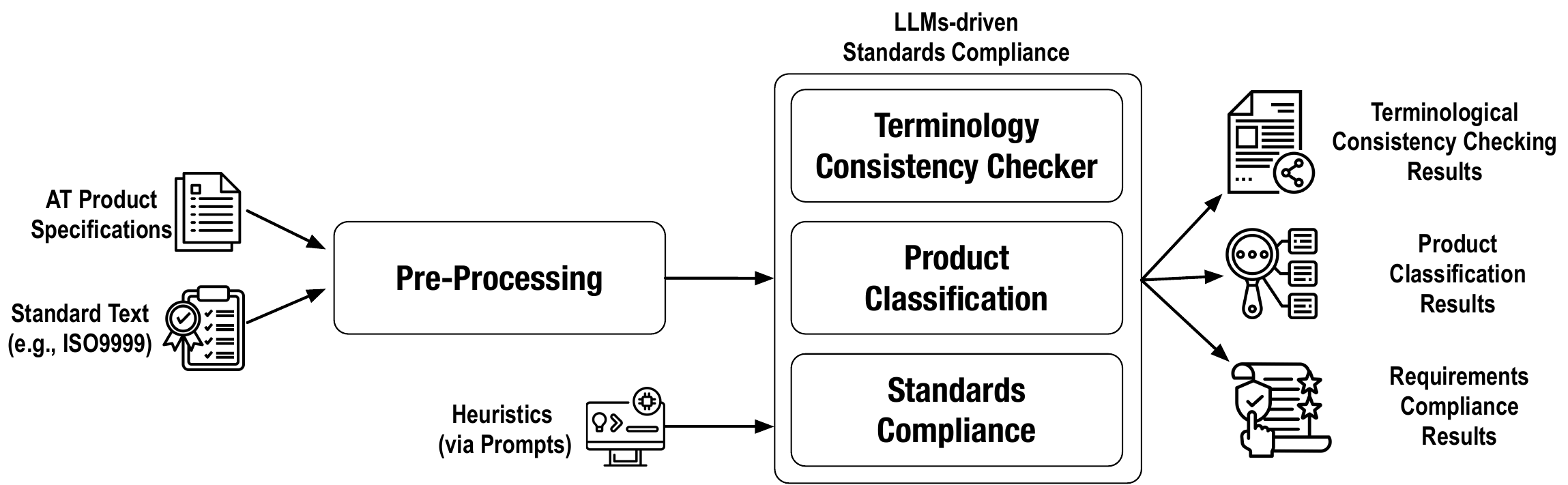}
\caption{\projname~Overview.}
\label{fig:overview}
\vspace{-1em}
\end{figure*}

\sectopic{Key Challenges.} Performing these tasks manually is time-consuming, error-prone and cumbersome. The diversity of various standards means that ensuring a new or modified AT product is compliant requires considerable effort. Changing standards and standards in new markets further complicate this analysis. It may be unclear how to describe many new AT devices, and inconsistent terminology in both standards and practices increases the difficulty. Many AT products often fail to adhere to relevant standard guidelines and undergo several rounds of compliance checks before being made available to end users~\cite{layton2024guidelines,bauer2012integrating}. We posit that \projname~will be helpful for both AT product manufacturers, which involve hardware and software products and the relevant national and international agencies responsible for ensuring that AT products comply with relevant standards.

\section{Proposed Framework}~\label{sec:approach}

Figure~\ref{fig:overview} shows an overview of \projname - our LLM-driven framework proposed for standards compliance of AT product specifications. \projname~focuses on three key steps in the AT standards compliance (discussed in Section~\ref{sec:introduction}): (1) checking the terminological consistency of product specifications against the terminology prescribed in the standards; (2) classifying the product(s) according to the correct classifications of the standards; and (3) checking the compliance of product specifications against the requirements of the correct product categories and other related standards. Below we explain the implementation plan for each task after discussing a common pre-processing step for all three tasks.

\subsection{Pre-Processing}~\label{subsec:preprocess}
The pre-processing step in \projname~focuses on analyzing the product specification ($SPEC$) and the standard text ($STAN$), and preparing the text for analysis for the subsequent three standards compliance tasks. For instance, we employ preliminary NLP pre-processing steps~\cite{sabetzadeh2024practical}, such as noise removal (i.e., stripping the two input documents of any extra or irrelevant information), tokenization, text chunking (to identify key noun phrases) and named-entity recognition (NER) for identifying the products or other categories of interest in $SPEC$ and $STAN$. For example, our Smart Knee example includes ``StrideTech-ProKnee'', ``StrideTech-ProKnee app", ``fall detector", ``smartphone", ``iPad" etc as NER entities. Note that pre-processing $STAN$ is a one-time event, and the results can be stored for any future analysis. 

\subsection{Terminology Consistency Checker}~\label{subsec:terminology}

For the terminology consistency checking task, we focus on identifying and extracting the main keywords from both the input documents based on the noun phrases identified earlier, according to our previous work on requirements glossary extraction and clustering~\cite{Arora:14a}. Once the keywords are identified, we use the NER results from the pre-processing step to select only the keywords of interest, such as product names or domain-specific keywords. After that, the keywords from $SPEC$ and $STAN$ are matched against each other to identify inconsistencies using string similarity matching (embeddings) and heuristics on the threshold for a match. The output of this task is the termnological consistency checking report for a given $SPEC$, which can be passed onto a human expert for review. For example, in our initial Smart Knee description, under ISO9999, it would be classified under orthotic and prosthetic devices related to lower limbs. 

After task 3, $SPEC$ would need to be further checked for adherence to ISO 8549 and ISO 10328 standards, including load testing and vocabulary usage. These could include ISO 8549-1: General terms for external limb prostheses and external orthoses -- use of definitions related to prosthetic devices, including what constitutes a knee prosthesis, the definition of a microprocessor-controlled prosthetic knee, and terms related to user safety and mobility; ISO 8549-2: Terms relating to external limb prostheses  -- definitions around support and stability, and context for the intended use and user needs that the StrideTech-ProKnee addresses.

\subsection{Product Classification}~\label{subsec:product_Classification}

We build a retrieval-augmented generation (RAG) approach~\cite{lewis2020retrieval} using large language models (LLMs) to implement this task. Product classification is a challenging task, as there could be hundreds of product categories in a given $STAN$, and it is difficult to build a classifier (e.g., a machine learning classifier) that can yield accurate results for multi-class classification with little training data and so many classes. For instance, StrideTech-ProKnee will be classified into more than one product category in ISO9999. We thus implement a hierarchical classification approach using RAG. Specifically, the process begins with the construction of a comprehensive knowledge base that includes detailed information from $STAN$, such as categories, definitions, and examples of previously classified products. This knowledge base is crucial for the retrieval component of the RAG model, enabling it to pull relevant information for product classification. The technical approach for hierarchical RAG-based classification and knowledge base will be additional key contributions of our research. 

For each new $SPEC$, the RAG model initiates a retrieval phase, querying the knowledge base to fetch relevant information based on semantic similarity, focusing on context and meaning rather than exact terminology matches. Utilizing the retrieved information, the generative component of LLMs predicts the product's classification, applying a hierarchical process to pinpoint the precise category by navigating from broader classifications down to specific subcategories. This helps manage the complexity of multiple categories and ensures accuracy. For example, for StrideTech-ProKnee specification, the RAG model will first classify the product in class 06 Orthoses and Prostheses and subsequently into other sub-categories, e.g., 06 24 33 Knee units. 

\subsection{Standards Compliance}~\label{subsec:ComplianceChecking}

The third task focuses on (i) identifying other relevant standards that the $SPEC$ should comply with based on the product classification results, and (ii) generating compliance recommendations for the current $STAN$. Currently, there is no systematic way of knowing which standards an AT product should comply with for a given country. As a part of \projname~development, we plan to develop a knowledge base to identify the standards relevant to a given product in a specific country. This is equivalent to the requirements traceability task for the actual compliance part, which has been widely studied in the RE literature. We build another RAG-based approach for tracing the key requirements in $SPEC$ to the $STAN$. The knowledge base in this task will greatly benefit all stakeholders in this research, as no such comprehensive databases exist.

In our StrideTech-ProKnee AT example, the RAG approach would generate trace links to the relevant testing processes and conditions in ISO 10328:2016, for each key product specification. This could include test methods for static, dynamic, and fatigue strengths, i.e., how the prosthetic knee withstands physical stresses and mimics natural gait patterns without failing; 
environmental testing -- subsections relevant to how prostheses must perform in various environments, including the sensor's role in monitoring the knee and surrounding environment in the StrideTech-ProKnee; and requirements for user safety and performance -- parts of the standard that address safety testing, especially related to the prevention of falls and ensuring stability.

Thereafter, we can also use the generative components of LLMs to provide a preliminary analysis of product specifications compliance with standard processes. The compliance rules will be drafted in natural language and inserted as prompts to LLMs. The output of this step is a requirements compliance results report on the other relevant standards, the traceability to the current $SPEC$, and compliance analysis. Engineers and acceptance testers can use these traceability documents to carry out analyses that all relevant standards and tests are being met by the device.


\section{Related Work}~\label{sec:related}

\subsection{AT Standards and Compliance}~\label{subsec:ATcompliance}
Standards are essential for ensuring a baseline level of quality, enabling clear communication, evaluating safety, and facilitating the comparison of products \cite{hobson2006standards,bougie2008iso}. Many countries have standards for AT products, and some follow the International Organization for Standards (ISO) for compliance purposes \cite{bernd2009existing,bauer2011assistive}. A number of studies have proposed better harmonisation of diverse AT standards~\cite{cooper1998harmonization}. 
Several others have emphasized the importance of three \projname~tasks proposed in this paper~\cite{sadeghi2018influence,elsaesser2019value,elsaesser2022standard}. For instance, Elsaesser et al.~\cite{elsaesser2022standard} emphasized that the lack of standard terminology in the AT domain leads to numerous communication issues and challenges for the globalization of AT products and services. Bauer et al.~\cite{bauer2011assistive} develop an AT device classification using ISO 9999, aiming to be consistent with various legislative acts. None of the previous works in AT research have leveraged advanced NLP techniques and attempted to build comprehensive knowledge bases to facilitate standards compliance.

\subsection{Requirements Engineering and Compliance}~\label{subsec:REcompliance}

Requirements compliance to regulations has been extensively studied in the RE literature due to the increasing complexity and rigor of regulations across various industries such as finance, healthcare, and assistive technology~\cite{akhigbe2019systematic,mustapha2020systematic,abualhaija2023legal}. Several studies in RE have introduced (semi-)automated approaches for identifying, interpreting, and integrating regulatory requirements into the software development lifecycle or checking their compliance against requirements~\cite{abualhaija2023legal}, with the compliance rules specified in various formats, e.g., natural language~\cite{kiyavitskaya2008automating}, templates~\cite{turetken2012capturing}, activity diagrams~\cite{soltana2018model}, formal notations~\cite{elgammal2016formalizing}, and semantic web-based methods~\cite{francesconi2023patterns}. The advent of NLP and ML technologies has opened new avenues for enhancing regulatory compliance in RE. Studies such as those by Cleland-Huang et al.~\cite{cleland:2010} and Guo et al.~\cite{Guo:17} have showcased the potential of NLP techniques in automating the traceability of requirements to regulatory documents, thereby reducing the manual effort involved and improving the accuracy of compliance checks. Additionally, research on the application of other AI techniques such as ML, question answering~\cite{abualhaija2022automated} or generative AI for predicting compliance issues and guiding the requirements elicitation process reflects an ongoing shift towards more intelligent and adaptive compliance management systems~\cite{arora2023advancing}. These advancements underscore a growing recognition of the need for innovative approaches to navigate the complexities of regulatory compliance, ensuring that software systems not only meet functional requirements but also adhere strictly to the regulatory frameworks governing their operation and use. Having said that, to the best of our knowledge, none of the previous works in RE have addressed the issue of standards compliance and specifically for AT product specifications.

\section{Conclusion}~\label{sec:conclusion}

This paper presents our vision of the \projname~framework to enhance compliance with standards in the assistive technology (AT) sector by applying advanced natural language processing techniques and large language models. By addressing three critical tasks—terminology consistency checking, product classification according to standards, and compliance checking against product categories and standards—\projname aims to streamline the complex and time-consuming process of ensuring that AT products meet the rigorous requirements set forth by national and international regulatory bodies. This framework aims to reduce the compliance burden on manufacturers, policymakers and national organizations responsible for managing AT compliance and significantly improve the safety, efficacy, and accessibility of assistive technologies for the diverse populations they serve.

As the next steps in this research, we are developing the prototypes for each task and the knowledge bases required for enabling product classification and standards compliance. We plan to thoroughly evaluate all tasks using different LLM configurations and AT standards. We further plan to actively engage with AT industry stakeholders, regulatory bodies, and AT researchers (including the third and fourth authors) to evaluate the outcomes of \projname. We posit that the \projname~vision presented in this paper lays the groundwork for a more efficient, reliable, and inclusive future for AT development and usage. Additionally, the lessons learned from this research would directly benefit RE in other domains, particularly regulatory compliance and the applications of LLMs in RE in practical contexts.

\bibliographystyle{IEEEtran}
\bibliography{paper.bib}
\end{document}